# Direct writing of high temperature superconducting Josephson junctions using a thermal scanning probe

*Ngoc My Hanh Duong,[1][*] Amanuel M. Berhane,[1] Dave Mitchell,[1] Rifat Ullah,[1] Ting Zhang,[1] He Zhu,[1] Jia Du,[1] Simon K.H. Lam,[1] Emma E. Mitchell,[1] and Avi Bendavid[1][*]*

[1] CSIRO Manufacturing, P.O. Box 218 Lindfield, NSW, Australia

* ngocmyhanh.duong@csiro.au, avi.bendavid@csiro.au

ABSTRACT

In this letter, we demonstrate for the first time the creation of Josephson-like superconducting nanojunctions using a thermal scanning probe to directly inscribe weak links into microstrips of $YBa_2Cu_3O_{7-\delta}$ (YBCO). Our method effectively reduces the critical current ($I_c$) over an order of magnitude. The resulting nanobridges exhibit clear evidence of Josephson effects, of SNS-type junctions, as shown by both the DC and AC Josephson effects. This approach provides a novel and flexible method for scaling up quantum mechanical circuits that operate at liquid nitrogen temperatures. Additionally, it offers a promising pathway for modifying properties of the junctions in-situ and post fabrication.

High-transition-temperature superconductors (HTSs) have garnered significant attention in both fundamental research and practical applications. The primary advantage of these HTS materials is their ability to become superconducting around liquid nitrogen temperature (~77

K), requiring simpler cryogenic systems for operation than low temperature superconductors (LTSs). Researchers have explored various methods to fabricate Josephson junctions (JJs) in these fascinating materials, including approaches based on grain boundaries or multilayer structures utilizing normal metals or semiconductors as barrier materials[1, 2]. An alternative approach is to define a nanoscale region within a superconducting microbridge to create a well-defined, non-superconducting barrier. However, fabricating these nanobridges from HTS materials remains challenging due to the short and anisotropic superconducting coherence lengths in these materials, typically about 2 nm in the a-b plane and 0.2 nm along the c axis. This makes the electrical properties of these JJs highly sensitive to chemical variations and structural defects at the atomic scale[3], which can occur during the fabrication process[4]. Common methods for fabricating HTS JJs include maskless focused ion beam etching and electron beam lithography. In many cases, a focused particle beam is used to directly write the JJs into the superconducting materials[5-7], while in conventional patterning processes, multiple lithography steps are required[8-10], which can degrade these materials. An alternative approach involves using an atomic force microscope (AFM)[11] to circumvent these issues; however, studies have shown unclear properties of the HTS JJs and require stringent nano-ploughing conditions[12].

Here, we propose a fast and fully controlled direct-write method to create nano-constrictions in pre-patterned YBCO microbridges using a thermal scanning probe. By adjusting the force and temperature of the heated tip, we can precisely define the nano-constriction on the superconducting microbridges. This technique operates under ambient atmosphere eliminating the need for vacuum. Standard four-point current-voltage (I−V) measurements performed at 77 K demonstrate a reduction of critical current ($I_c$) by over an order of magnitude. Our technique demonstrates the capability to directly create nanoscale Josephson junctions, marking progress toward fabricating planar YBCO films for HTS circuit-based devices.

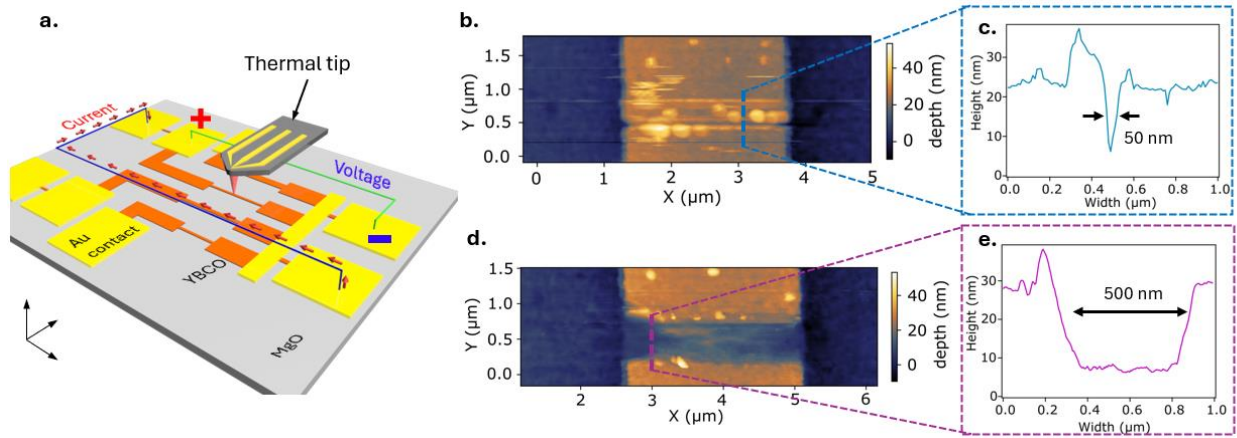

**Figure 1.** a) Schematic diagram of the microstrips and electrical measurement setup. b, d) AFM scans using the thermal probe of two different cuts. c, e) Profiles of the cuts shown in b and d, respectively.

Initially, the microbridges were patterned into the YBCO thin film, which was deposited on a 10 mm x 10 mm MgO (100) substrate, using photolithography and argon ion beam milling technique. This fabrication process resulted in devices consisting of 32 superconducting strips, with widths ranging from 2 to 8 μm, as illustrated using the schematic in figure 1a.

The samples were then loaded into a Nanofrazor, where an ultrasharp, heatable Si tip with a radius of less than 2 nm and 750 nm length was used to directly scribe across the 2 μm-wide YBCO microbridges, which were 30 nm thick. The advantage of the Nanofrazor is that it allows for imaging of the surface before, during, and after patterning, similar to an atomic force microscope[13]. After trial runs on several YBCO microbridges, we determined that the optimal scribing conditions were a combination of temperature of 1350°C and a forward height of 0 nm, enabling us to cut through the 30 nm thick YBCO microbridges in a single pass of a duration less than 60 seconds (tip velocity is around 0.1 μm/s), as shown in figures 1b and 1d. This method is more efficient than typical AFM techniques[12, 14], which require either a diamond-like coated tip or several hundred cycles of back-and-forth AFM tip movement to create a single structure. Our current understanding is that the combination of heat and force is crucial for cutting YBCO or other materials, which differentiates our technique to conventional AFM techniques. Due to the small tip diameter, the tip can achieve pressures of up to 4 GPa[15]. The applied heat lowers the material's breaking strength, allowing the tip to reach the necessary pressure for effective cutting of YBCO[16]. The dwelling time for one pixel is 40 µs, which offers enough time for the heat to penetrate the YBCO layer of 30 nm thick (see Supporting Information, Fig. S2 for our modelling of heat transfer in YBCO). Another advantage of this technique is the ease of controlling the depth of the cut through the closed-loop patterning process incorporated into the Nanofrazor, which provides simultaneously imaging of the pattern. The linewidth can be controlled by setting parameters (as shown in figures 1c and 1e, with cutting lengths of 50 and 500 nm, respectively). The minimum linewidth, determined by the tip radius, is 15 nm, which is three times better than other AFM studies[12, 17].

To investigate the properties of the nanobridge created by thermal scribing, we gradually decreased the width of the nano-constrictions (W) by trimming from the right to the left side of the bridge and measuring the critical current of the structures after cutting each time, as shown schematically in figure 2a. Two 2-µm-wide YBCO microbridges, with a thickness of 113 nm and an initial critical current density ($J_c$) of 2.4 MA.cm$^{-2}$, were aligned perpendicular to the tip scan axes and scanned in imaging mode using the Nanofrazor. Subsequently, the tip was lowered into contact with the microstrip at the desired position for creating the nano-constrictions, directly scribing onto YBCO with 2 nm pixel steps and a 40 µs tip contact time per pixel. During the thermal scribing process, the probe was heated and maintained at a writing temperature of $T_w$ = 1350°C using an on-tip resistive heater and monitored by an on-tip thermometer.

The current-voltage characteristics (IVC) of the bridges were measured using standard 4-terminal methods on a custom-built probe within a dip probe system shielded with mu-metal and cooled with liquid nitrogen to 77K (see more details in Methods). Figure 2b presents the initial I-V curves of the two 2 μm-wide bridges in the same block used in the experiment. Bridge 1 exhibited an $I_c$ of 4.9 mA and a normal resistance ($R_n$) of 619 Ω, while bridge 2 showed an $I_c$ of 5.4 mA and $R_n$ of 693 Ω. It is noted that these $R_n$ values correspond to the resistive state of the microbridges in a high voltage range of up to a few volts. When the remaining width of bridge 1 was reduced to ~200 nm and the length is ~300 nm, the $I_c$ decreased to ~0.3 mA, over an order of magnitude drop from the initial $I_c$ of 4.9 mA (see Fig. 2c). This demonstrates the potential of using this technique to actively modify the $I_c$ of nanojunctions in existing superconducting circuits without the need of multiple lithography steps. The $I_c$ values decreased as a function of the bridge width and followed the calculated values assuming the $J_c$ of the film remained unchanged, indicating that the technique is highly localized and not damaging the surrounding areas. The normal resistance ($R_n$) of the microbridges as a function of bridge width (see Fig. 2d) remained unchanged, further supporting the localization of the thermal scribing technique. The effective width ($w_e$) of these two nanobridges, however, are not resolved completely as there is residual YBCO on the edge.

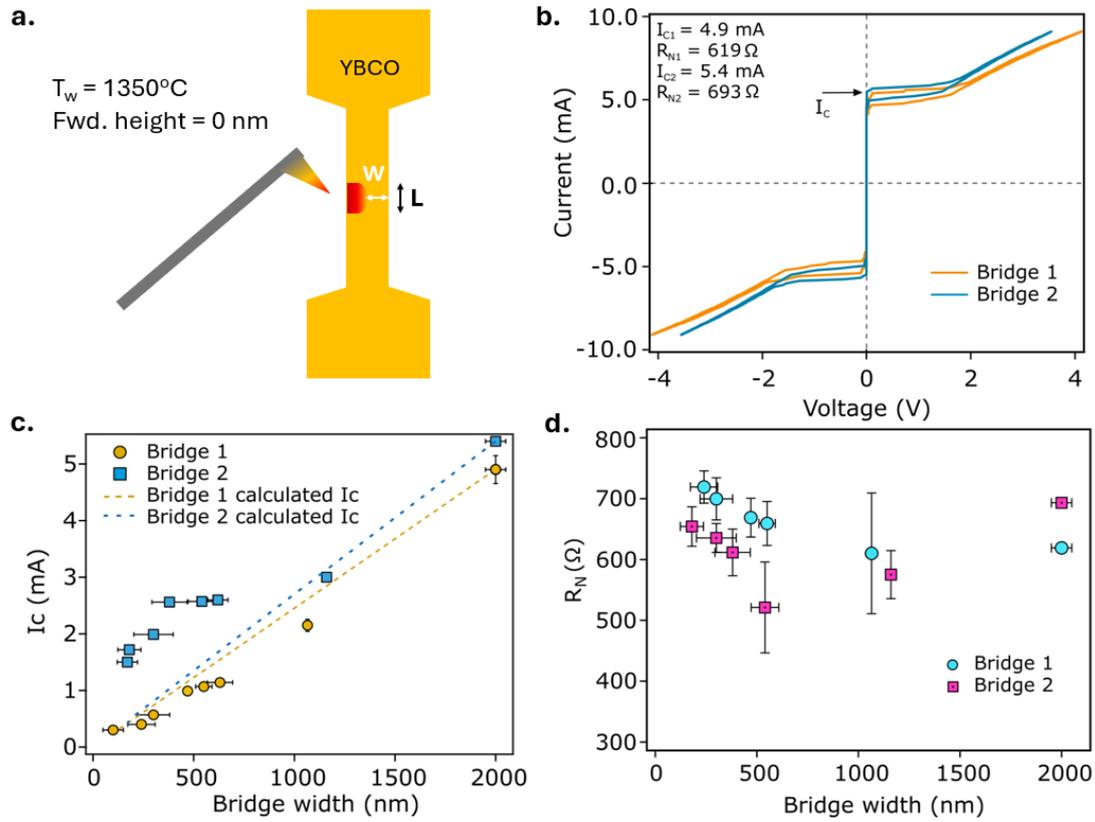

**Figure 2.** a) Schematic diagram of the thermal scribing process on YBCO microbridges, where w represents the bridge width and L denotes the bridge length. b) Initial current-voltage (I-V) curves from two microbridges in the study, showing critical currents $I_c$ of 4.9 mA and 5.4 mA, respectively with the slope of the I-V curves giving normal resistances $R_n$ of 619 Ω and 693 Ω, respectively. c) Critical current $I_c$ of bridges 1 and 2 as a function of bridge width (yellow circles and blue squares), along with calculated $I_c$ values assuming a uniform $J_c(x)$ along the bridges. d) Normal resistance $R_n$ at high bias currents for bridges 1 and 2 (blue circles and pink squares), proving the localization effect of the thermal scribing technique. All the electrical measurements were performed at 77K.

To better understand the Josephson behaviours and investigate the effective width of our nano-constrictions, we measured the dependence of $I_c$ of bridge 1 on an external magnetic field ($B_a$). When a field is applied perpendicular to the plane of the junction of width w, a uniform junction

($J_c(x)$ = constant across w) would show the Fraunhofer behaviour, which is given by $I_c = I_{c0}$ |sin k/k|, where $k = \frac{\pi B_a h_e w}{\Phi_0}$, $h_e$ is effective thickness of the barrier, and $\Phi_0$ is the flux density. At 70 K, we obtained a Fraunhofer-like diffraction pattern for bridge 1 (see the inset in figure 2a for AFM scan of nanobridge 1). The pattern is reasonably symmetric around $B_a = 0$, indicating the presence of the DC Josephson effect. However, this is not an ideal Fraunhofer pattern, as evidenced by non-zero minima and non-uniform spacing of the minima, which suggests a non-uniform $J_c(x)$ along the junction width. The magnetic field response of our nanojunction between two superconducting electrodes yields a periodicity of ~27 mT. This periodicity aligns well with an analytical model[18], which accounts for flux focusing of the flat electrodes, significantly altering the period such that $\Delta B_0 = 1.84\Phi_0/w^2$. This model predicts a junction width of ~200 nm for a periodicity of ~27 mT, which is in good agreement with our expected remaining width of ~ 200 nm of the nanobridge 1 after cutting.

The temperature dependence of $I_c$ and $R_n$ of the nanobridge is shown in figure 2b. The critical current $I_c$ scales with $(1 - T/T_c)^\alpha$, yielding an α value of 1.3 after fitting. For short nanobridges, the exponent α is expected to be in the range of 1-1.5, whereas for long bridges, it is typically 2-2.5[19-21]. Our nanobridge length of approximately 300 nm falls within the regime of short nanobridges. The reduction in $R_d$ as the temperature decreases suggests that the weak link acts as either a normal conductor or a poor superconductor, indicating the formation of superconductor-normal-superconductor (SNS) or superconductor-poor superconductor-superconductor (SS'S)-type junctions. The resistance (~20 Ω) is approximately 200 times larger than typical nanolithography and ion-irradiated weak links[22] and two orders of magnitude higher than the recent direct write SNS junctions using helium ion irradiation[5].

Next, we study the AC Josephson effect in nanobridge 1. The AC Josephson effect established the following voltage-frequency relationship[23]:

$$f = \left(\frac{2e}{h}\right)V_0 , \frac{2e}{h} = 0.4836\frac{GHz}{\mu V} \quad (1)$$

Where $V_0$ is the DC voltage across the junction when biased above its critical current $I_0$. Given that $\frac{2e}{h} = 0.4836\frac{GHz}{\mu V}$, for $V_0$ ranging from $\mu V$ to $mV$, the frequency $f$ spans microwave, mm, and sub-mm or THz bands. Under RF radiation, ladder-like voltage steps, known as Shapiro steps, form and are determined by the frequency-voltage relationship. Figure 3c shows the microwave responses of nanobridge 1 at frequencies of 19.6 GHz and 19.8 GHz, while Figure 3d presents I-V curves of nanobridge 1 under microwave radiation (19.8 GHz) at various power levels at 77K. Measurements at 50 GHz were unsuccessful due to inefficient RF coupling, caused by a mismatch in frequency transmission, which failed to suppress $I_c$. Effective RF power coupling to the nanobridge was achieved only at approximately 20 GHz.

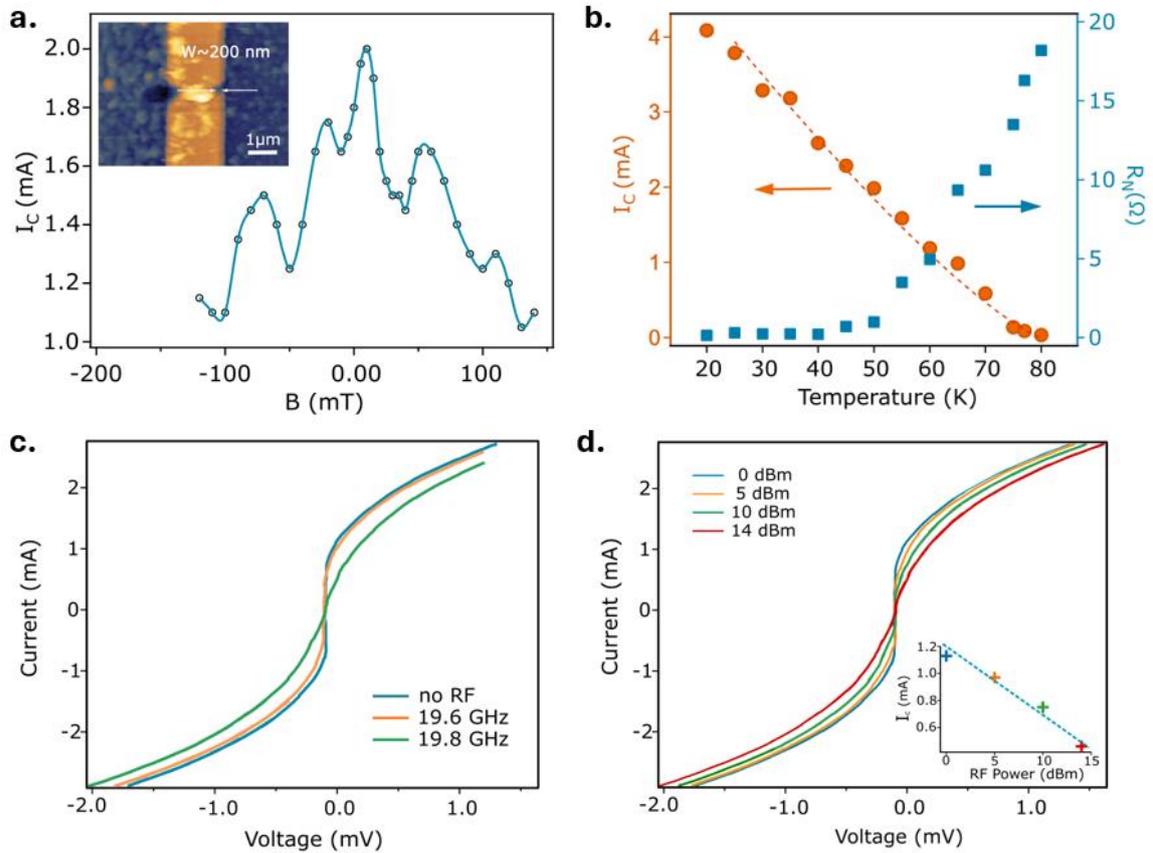

**Figure 3.** DC and AC Josephson effect in nanobridge 1. a) The effect of external magnetic field on the critical current of the bridge 1, the inset is an AFM scan of nanobridge 1 performed by using thermal scanning probe. The field is applied perpendicular to the plane of the bridge revealing a Fraunhofer-like pattern at 70 K. b) Critical current versus temperature (orange cycles) of nanobridge 1 with exponential fits (orange, dashed line), and $R_n$ as a function of temperature (blue squares). c) Measured dc I–V behaviours of the junction with and without microwave radiation of different frequencies at 77 K. d) I-V curves of a Josephson junction under microwave radiation (19.8 GHz) with different RF power levels. The inset shows the $I_c$ as a function of RF power.

Although the suppression of $I_c$ at zero voltage was clearly observed under microwave excitation, Shapiro steps beyond the zeroth order could not be resolved. Several factors could explain this observation. First, the critical current ($I_c$) remains relatively large, and the thermal rounding effect may smear the microwave-induced Shapiro steps, particularly if the step heights are small. Second, the junctions are operating in a low-frequency regime, as indicated by the normalized frequency $\Omega = \frac{f_s}{f_c} = \frac{f_s}{\frac{2e}{h}I_c R_N} < 0.1$, where $f_s$ is the RF signal frequency and $f_c = \frac{2e}{h}I_c R_n$ is the junction's characteristic frequency. A simulation by Russer[24] shows that at small values of $\Omega$, the height of the Shapiro steps becomes increasingly smaller, and in the limit as $\Omega \to 0$, the steps disappear and only the maximum current $I_c$ at zero voltage depends on the incident microwave power. Indeed, a linear decrease of the $I_c$ with an increase in of the microwave power is observed for our nanobridge indicating the presence of a Josephson component in the supercurrent (see inset in Fig. 3d). It is also worth noting that the resolution of the IVC data collected using the Pico Scope software was insufficient to distinguish the very small differences in $I_c$, as illustrated in figures 3c and 3d. Improving the measurement techniques will be considered in future work.

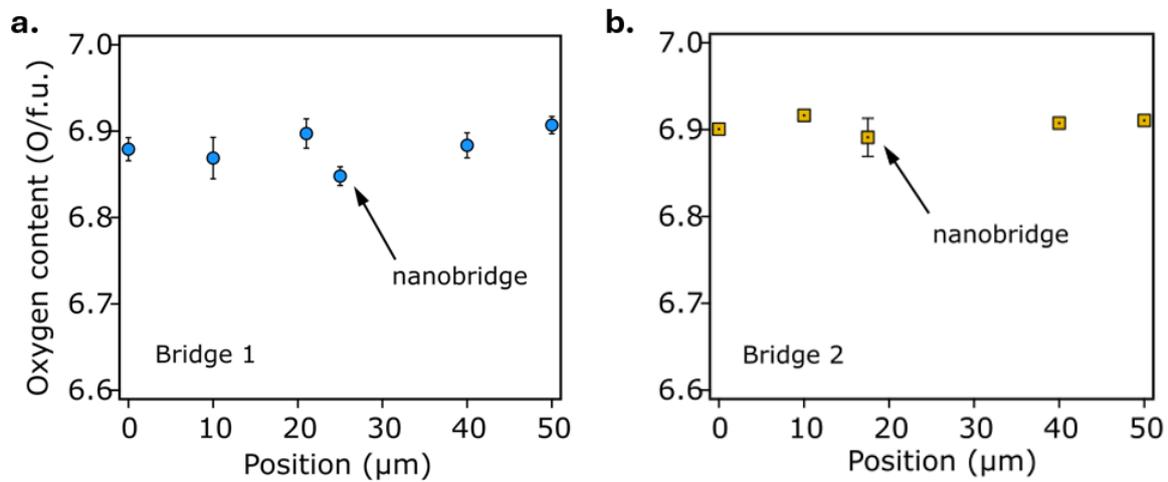

**Figure 4.** Oxygen stoichiometry analysis of bridge 1 (a) and bridge 2 (b) extracted from Raman spectra. Error bars are calculated using the standard deviation of 3-5 repeated measurements at each location.

To confirm the localization of this technique, we further characterized the microbridges after writing using Raman spectroscopy. Figures 4a and 4b show the oxygen stoichiometry calculated from Raman spectra for two of the microbridges used in the experiment. The calculation is performed based on the correlation between the peak position of the O(4)-Ag phonon mode and the oxygen deficiency within the YBCO crystal structure where a blue-shift in the position of this peak represents an increase in oxygen stoichiometry[25]. Positions along the length of the YBCO microbridge were chosen for investigation by Raman spectroscopy to characterise the relative oxygen content present to assess the damage done to the bridge by the thermal probe. Along the bridge, we observed a consistent oxygen concentration within the crystal lattice, averaging around 6.9 mol per formula unit (f.u.), indicating the material remains undamaged. The spectra from the nano-constriction areas, show a slight reduction in oxygen stoichiometry, suggesting a possibility of oxygen depletion happened in these regions. Despite this trivial change, the areas remain within the superconducting range, specifically above a mole ratio of 6.2 O/ f.u.[26] when the YBCO becomes insulating (see Supporting Information, Fig. S1), further confirming the localization of the technique.

In conclusion, we have introduced a one-step lithography technique that allows for the direct writing of Josephson junctions from high-temperature superconductor YBCO for the first time using a thermomechanical etching technique. The nanobridge obtained show evidence of Josephson effects in an SNS or SS'S-type JJs. The technique offers several advantages, including the simplicity of fabricating junctions and the flexibility to integrate them in desired locations on pre-patterned circuits without the need for further complex lithography processes. Additionally, it allows for the direct trimming of pre-existing junctions to modify critical current, which can enhance the performance of superconducting quantum interference devices (SQUIDs). In the future, the use of parallel writing with up to 10 tips in a single write might enable the simultaneous writing of multiple nanojunctions, paving the way for scaling up devices. Given the localization of this technique, it holds potential for extending its application to other high-temperature superconductors that are sensitive to disorders and chemicals, where traditional lithography methods may be challenging and prone to damage the junctions during fabrication.

**Materials and Methods**

**Pre-patterning of the YBCO microbridges**

The YBCO microbridges were pre-patterned using photolithography and Ar ion beam etching techniques. A 113 nm-thick epitaxial c-axis YBCO film with a 50 nm in situ Au film was deposited on the MgO substrate by Ceraco, GmbH using e-beam evaporation. The Au/YBCO film was then patterned and etched to form the microbridges. The in-situ Au was removed from the junction area using an Ar-sputtering technique. A second lithography step was conducted to make Au contact pads of ~ 300 nm thick.

**Thermal scribing of pre-patterning YBCO microbridges**

Patterning of YBCO microbridges is performed using a commercial NanoFrazor t-SPL tool (Heidelberg Instruments), which utilizes a nanoscale probe to locally heat and directly trim YBCO with nanoscopic resolution. The software offers precise control over the size of the cut, including width, length, and depth. The temperature is adjustable within the range of 300-1400°C, and the resolution of the cut is approximately 15 nm.

**Transport measurements**

IVC were measured using the standard four-terminal method at 77 K. The superconducting bridges mounted on a dip-stick probe that was immersed into a liquid nitrogen dewar positioned inside a three-layer mu-metal shield. An Advantest TR6143 source measure unit was used as a current source and a Hewlett-Packard 3458A multimeter was used to measure the voltage across the superconducting bridge. Temperature dependence critical current and Fraunhofer diffraction pattern measurements were carried out in low fields up to 140 mT in a Physical Property Measurement System (PPMS, Cryogenic Ltd).

**RF measurement setup**

For the cryogen-free cooling process, the sample was housed in a copper enclosure and mounted onto the cold head of a two-stage pulse-tube cryocooler (Janis Research Inc., model PTSHI-SRP-062B). Each DC line connected the device contact pads to a printed circuit board (PCB), which featured a simple filter with a 1 kΩ resistor and a 1.5 nF capacitor in series between the DC line and the ground plane. A battery-operated current source applied the current, and the I-V characteristics were captured using Pico Scope software. For the RF measurements, a coaxial cable was connected to the sample through an SMA connector, and an RF signal with a frequency of 20-50 GHz was generated using an Agilent Technologies

E8257D signal generator. The microwaves were delivered directly to the nanobridges via the coaxial cable.

**Raman spectroscopy**

Raman spectroscopy measurements were performed using a 514 nm laser at room temperature with a Renishaw InVia Reflex spectrometer equipped with a Peltier-cooled charged coupled device detector. The Raman spectra were recorded in a backscattering geometry and the laser power was set to 50 mW, using a 50x objective (NA = 0.75), resulting in a nominal spot size of approximately 1 μm. The laser spot was used to probe local structure and oxygen content within the YBCO film along the microbridge.

ACKNOWLEDGMENT

The authors thank Jeina Y. Lazar, and Roger Chai for supporting in the fabrication of devices. We also thank Dr. Phil Martin for fruitful discussions.